\documentclass[twoside,slac_one]{revtex4}
\usepackage{graphicx}
\usepackage{fancyhdr}
\usepackage{amsmath} % American Mathematics Society standards
\usepackage{bm}% bold math
\usepackage{amsxtra}
\usepackage{amssymb}
\usepackage{amsthm}
\usepackage{latexsym}
\usepackage{lscape}
\usepackage{multirow}

\newcommand{\powheg}{{\sc{powheg}}}
\newcommand{\pythia}{{\sc{pythia}}}
\newcommand{\fewz}{{\sc{fewz}}}
\newcommand{\mcfm}{{\sc{mcfm}}}
\newcommand{\mstw}{{\sc{mstw08}}}
\newcommand{\pdflhc}{{\sc{pdf4lhc}}}
\newcommand{\geant}{{\sc{geant}}}
\newcommand{\met}{$\rlap{\kern0.25em/}E_T$~}

\begin{document}

%Title of paper
\title{W and Z Boson Cross Section and W Asymmetry at CMS}

% Repeat the \author .. \affiliation  etc. as needed
%
% \affiliation command applies to all authors since the last
% \affiliation command. The \affiliation command should follow the
% other information

\author{Jiyeon Han}
\affiliation{Department of Physics and Astronomy, University of Rochester, Rochester, NY, USA}

\begin{abstract}
We report the measurement of the rates and asymmetries of inclusive and differential production
 of W and Z vector bosons in pp collisions at $\sqrt{s}=7$ TeV.
The data consists of 36 $pb^{-1}$ collected in the Compact Muon Solenoid (CMS) detector
 at the Large Hadron Collider (LHC).
The measured inclusive cross sections are $\sigma(pp\to WX)\times B(W\to\ell\nu) = 10.31\pm 0.02$(stat.)$\pm0.09$
(syst.)$\pm 0.10$(th.)$\pm 0.41$(lumi.) nb and $\sigma(pp\to ZX)\times B(Z\to\ell\ell) = 0.975\pm 0.007$(stat.)$\pm0.007$
(syst.)$\pm 0.018$(th.)$\pm 0.039$(lumi.) nb.
The measured inclusive cross sections and also its ratio of W to Z or $W^{+}$ to $W^{-}$ agree with
 NNLO QCD cross section calculations and current parton distribution functions (PDFs).
The differential production asymmetry of W boson as a function of the lepton pseudorapidity
 in the final state is also measured and compared with various PDFs. 
\end{abstract}

%\maketitle must follow title, authors, abstract
\maketitle

\thispagestyle{fancy}

% body of paper here - Use proper section commands
% References should be done using the \cite, \ref, and \label commands
% Put \label in argument of \section for cross-referencing
%\section{\label{}}

%%%%%%%%%%%%%%%%%%%%%%%%%%%%%%%%%%
\section{Introduction}
\begin{figure}[ht]
\centering
\includegraphics[width=70mm]{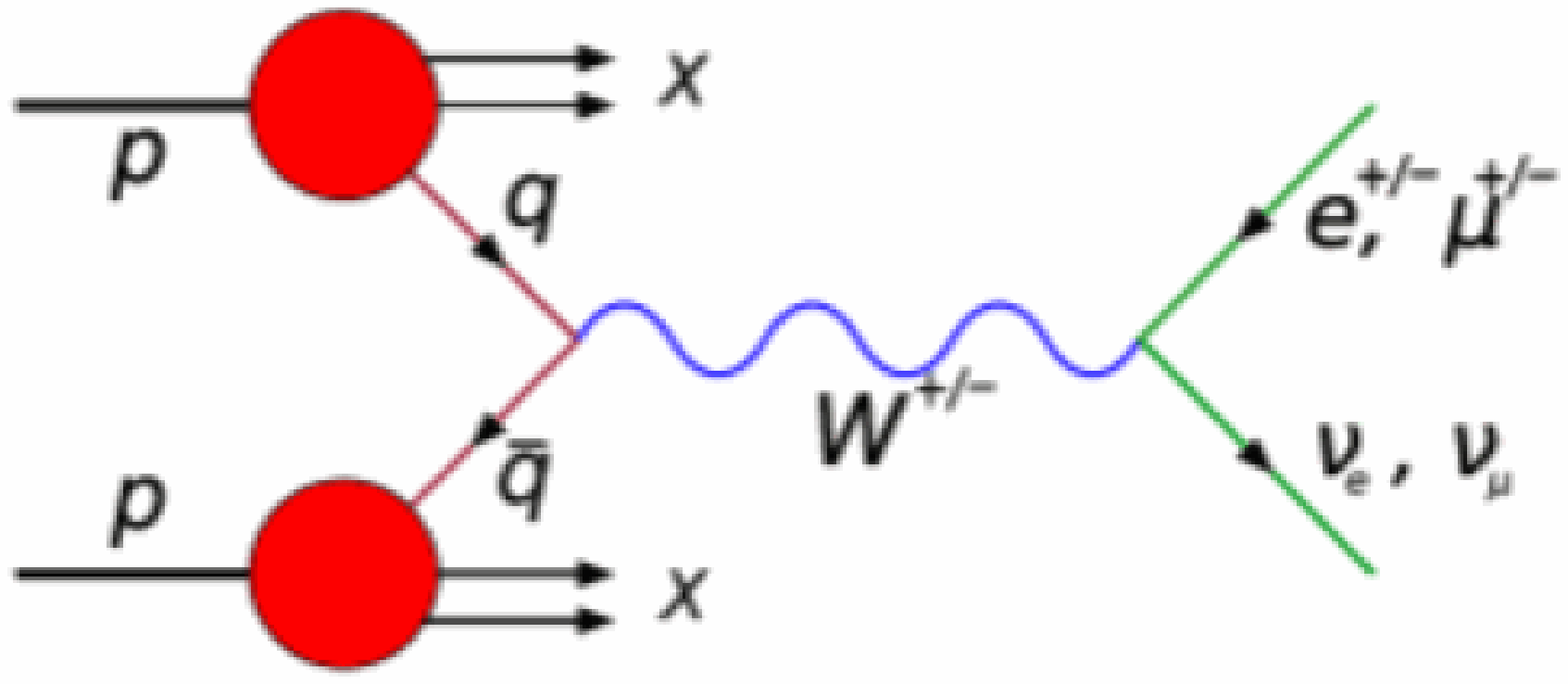}
\includegraphics[width=67mm]{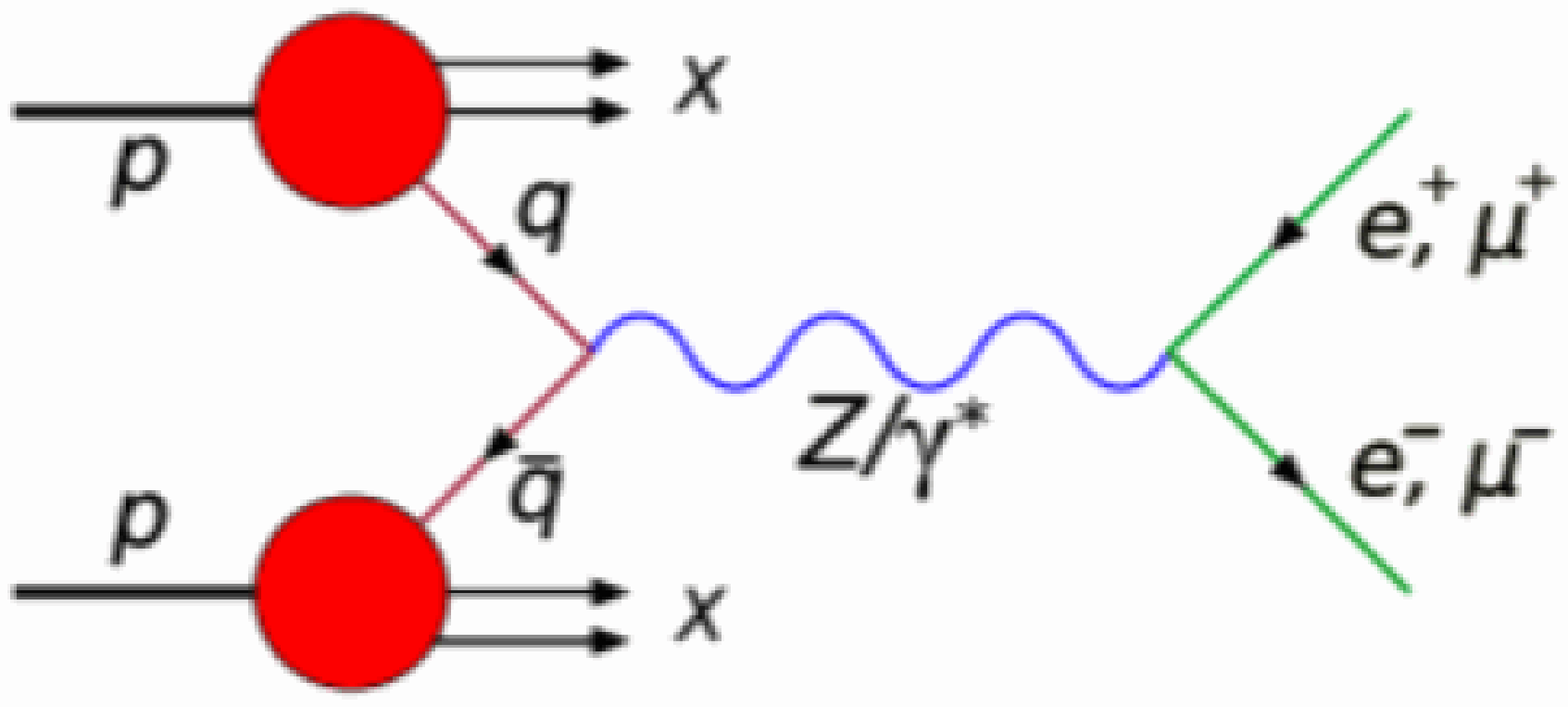}
\caption{W and Z boson production diagrams at LHC (pp collision).} \label{fig:wz}
\end{figure}

The inclusive cross section of any physics process is determined by Born-cross section ($\hat{\sigma}$)
 and the parton distribution functions (PDFs).
 Therefore, the measurement of the inclusive cross section tests calculations
 based on the higher order perturbative Quantum Chromodynamics (QCD) and also PDFs.
Especially, W and Z boson processes shown in Figure \ref{fig:wz} are well understood
 and unique signature with high rate, which give a useful information of QCD and PDFs.
Moreover, these processes are a good tool to calibrate the detector
 in the early stage and the total cross section of these processes can be used as the luminosity
 candles at LHC. W or Z process mediates for many complex final states like top or new physics,
 so the measurement of W and Z production process paves the way for understanding top or new physics processes.

In the W production mechanism, it is polarized due to the parity violation at production and it gives a strong
 asymmetry in the lepton decay. This asymmetry measurement constrains for the ratio of u and d quark,
 particulary on the sea quark contribution at LHC. Therefore, the asymmetry measurement of W boson production
 is important input of the global PDFs fit.

Here, we measure the inclusive cross section and its ratio of W and Z production decaying to electrons 
 or muons with the data sample taken from 2010 LHC operation period (36 $pb^{-1}$).

%%%%%%%%%%%%%%%%%%%%%%%%%%%%%%%%%%
\section{The CMS Detector}
A detailed description of the CMS detector can be found in Ref. \cite{cmsdet}.
The CMS detector is a superconducting solenoid, of 13 m in length and 6 m in diameter,
 which provides an axial magnetic field of 3.8 T.
The innermost layer is a silicon tracker which covers the pseudorapidity, $|\eta|<2.4$ ($\eta = -ln~\tan(\theta/2)$
 where $\theta$ is the polar angle of the trajectory of the particle with respect to the beam direction.
Outside of the silicon tracker, the scintillating crystal electromagnetic calorimeter (ECAL)
 and the brass/scintillator hadron calorimeter (HCAL) follow.
The ECAL provide coverage of $|\eta|<1.479$ in the barrel region and $1.479 < |\eta| < 3.0$
 in the endcap region.
The muon system is made of three technologies, drift tubes, cathode stript chambers, and 
resistive plate chambers. The muon detector covers $\|eta|<2.4$ region and has about 2 $\%$ 
 transverse momentum resolution in the relevant muon $p_{\rm T}$ range with the silicon track
 matching.  

%%%%%%%%%%%%%%%%%%%%%%%%%%%%%%%%%%
\section{Data Set and Event Selection}
The data sample consists of 36 $pb^{-1}$ collected by CMS detector at LHC during 2010.
The analysis is performed for the electron and muon channel for W/Z production.
The events from the high $E_{T}$ electron or the single high $p_{\rm T}$ muon trigger
 is used to select the sample.
In the muon channel, the muon is selected to have $p_{\rm T}>25$ GeV for W boson
 and $p_{\rm T}>20$ GeV for Z boson in $|\eta_{\mu}|<2.1$.
The good track quality selection is required to find a good muon
 and the impact parameter cut, $d_{xy}<2$ mm, is also required to reduce the cosmic muon
 background. 
The electron is required to have $E_{T}>20$ GeV for W and Z boson in a good detector fiducial
 region, $|\eta_{e}|<1.44$ and $1.57<|\eta_{e}|<2.5$.
The high $E_{T}$ electron cluster is matched to a high $p_{\rm T}$ track and also
 required to pass the electron identification (ID) variables like a shower shape and the conversion reject
 variables. Both electron and muon are required to be the isolated object.
Since the W production has large Drell-Yan background contamination, the Drell-Yan background is reduced
 by rejecting the event with the second isolated lepton with loose selection.

%%%%%%%%%%%%%%%%%%%%%%%%%%%%%%%%%%
\section{Acceptance and Efficiency}
\subsection{Acceptance}
The acceptance is calculated using \powheg~Monte Carlo (MC) \cite{powheg1,powheg2,powheg3} generator interfaced with \pythia ~parton showering.
The \powheg~generator is combined with a \geant~\cite{geant} simulation of the CMS detector.
The acceptance is the event fraction of the event with kinematic selection ($E_{T}(e),~p_{\rm T}(\mu)>20,~25$ GeV
for W selection, and $E_{T}(e),~p_{\rm T}(\mu)>20$ GeV for Z selection in the detector fiducial region)
to the generated events.
The acceptance for the Z boson measurement is restricted by the Z mass range, $60<M_{\ell\ell}<120$ GeV.
The total acceptance is 50 $\%$ for W boson and 40 $\%$ level for Z boson.
Table \ref{tab:acp} summarizes the acceptance for each channel.

\begin{table}[ht]
  \begin{center}
    \caption{The acceptance ($A$) of W or Z boson production.}
    \begin{tabular}{|l|c|c|}  \hline
%      \multirow{2}{*}{Process} & \multicolumn{2}{c|}{$A_{W,Z}$} \hline \\ 
%      & $\ell = e $   & $\ell = \mu$  \\ \hline
      Process                  & $A$ ($\ell = e $) & $A$ ($\ell = \mu $)   \\ \hline 
      $W^{+} \to \ell^{+}\nu$  & $0.5017 \pm 0.0004$ & $0.4594 \pm 0.0004$ \\ 
      $W^{-} \to \ell^{-}\nu$  & $0.4808 \pm 0.0004$ & $0.4471 \pm 0.0004$ \\ 
      $W \to \ell\nu$          & $0.4933 \pm 0.0003$ & $0.4543 \pm 0.0003$ \\  \hline
      $Z \to \ell^{+}\ell^{-}$ & $0.3876 \pm 0.0005$ & $0.3978 \pm 0.0005$ \\ \hline
    \end{tabular}
    \label{tab:acp}
  \end{center}
\end{table}  

\subsection{Efficiency}
The efficiency is determined for the offline reconstruction of the lepton,
 the lepton selection with the lepton identification (ID) and isolation criteria, and the trigger efficiency.
The efficiencies are estimated using the tag and probe method using Z sample.
The tagged leg has a tight lepton selection to clean up the sample and the probe leg is used
 to determine the efficiency.
The remaining background in the sample is subtracted using Z mass fitting method.
The efficiencies are estimated in the data and the simulation sample (MC) and the efficiency 
 scale factor of data to MC is applied into MC to correct the efficiency in MC.
Table \ref{tab:effi} shows the total efficiency for W and Z process in the electron and muon channel.

\begin{table}[ht]
  \begin{center}
    \caption{The total efficiency of W or Z boson production.}
    \begin{tabular}{|l|c|c|} \hline
      Lepton channel & W process  & Z process \\ \hline
      Electron  & $0.735\pm 0.009$ & $0.609\pm 0.011$ \\
      Muon      & $0.848\pm 0.008$ & $0.872\pm 0.002$ \\ \hline
    \end{tabular}
    \label{tab:effi}
  \end{center}
\end{table}  

\section{Background Estimation}
\subsection{Background and Signal Extraction for W Production}
\begin{figure}[ht]
\centering
\includegraphics[width=80mm]{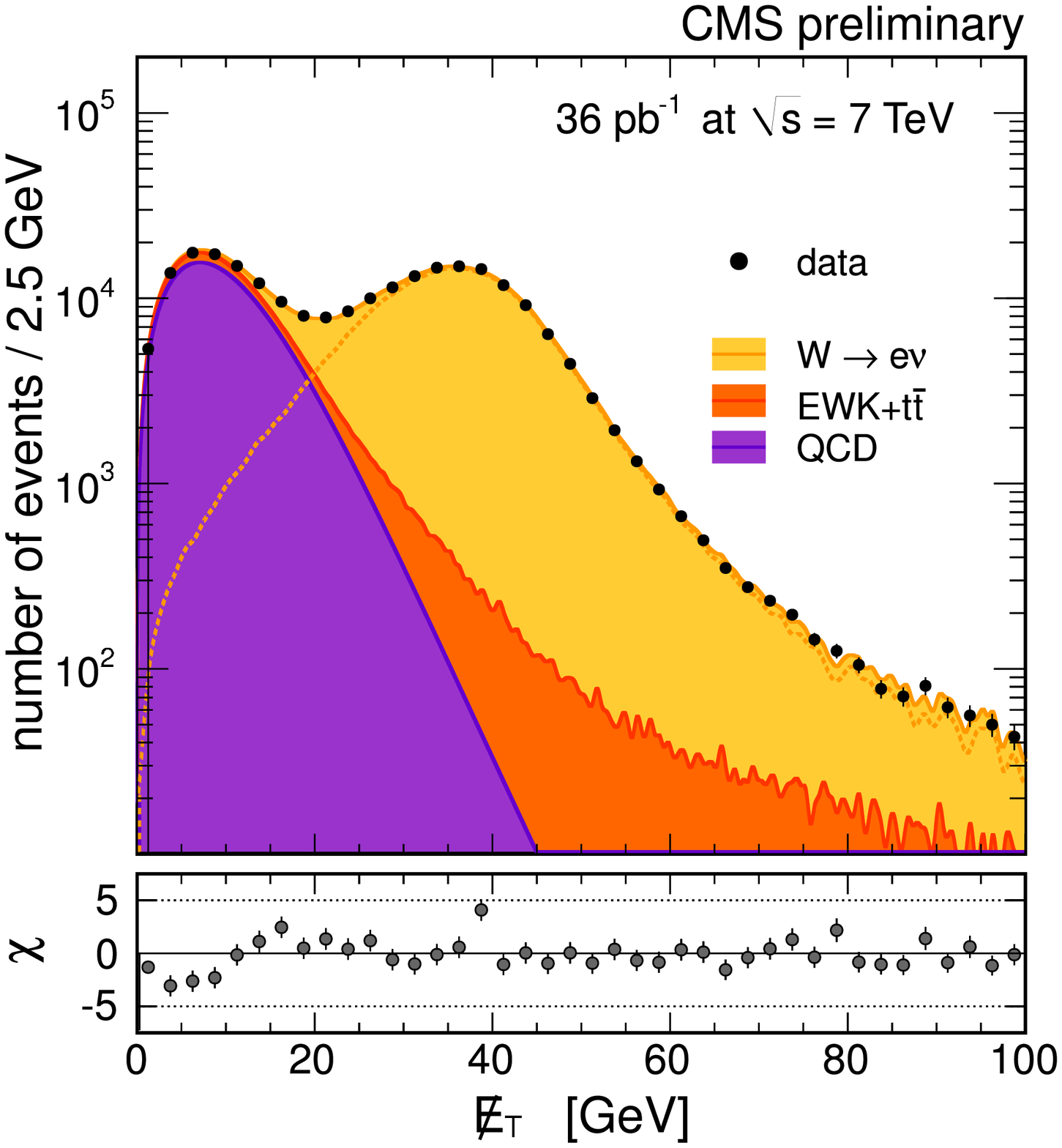}
\includegraphics[width=80mm]{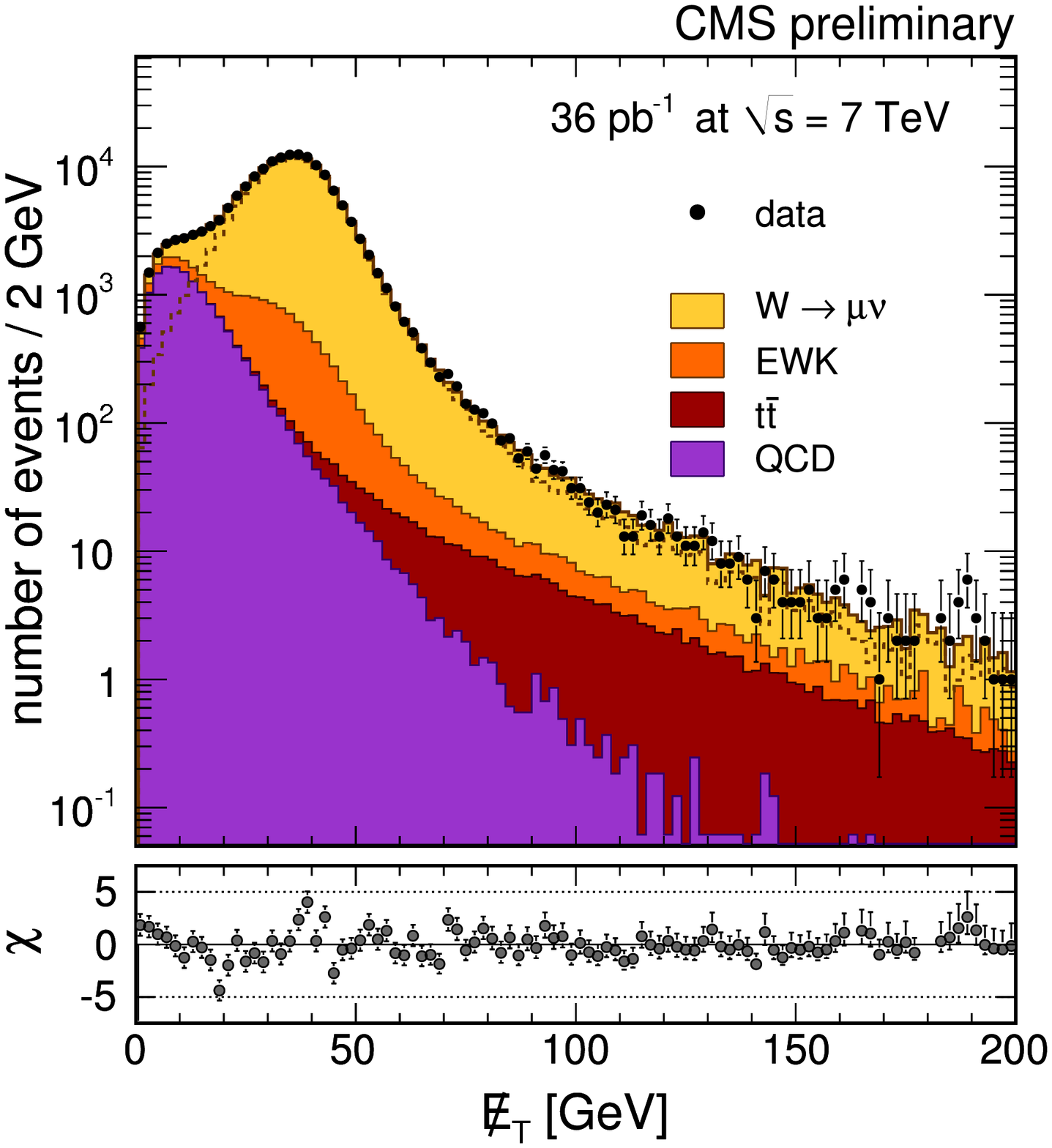}
\caption{\met distribution of $W\to e\nu$ and $W\to \mu\nu$ with the background contribution.} \label{fig:w_met}
\end{figure}

 The main background source in W production is QCD multi-jet and Drell-Yan process.
QCD background is estimated from the data directly using \met spectrum fit method.
The background shape of \met is obtained from QCD enriched sample and the signal shape
 is obtained using MC and $Z\to \ell^{+}\ell^{-}$ data for the hadron recoil tuning.
The background estimation from \met fit is also confirmed by two independent methods,
 $M_{T}$ shape fit and Isolation energy vs. \met method (ABCD method).
The QCD background rates estimated from these different methods are consistent each other
 within the statistical uncertainty.
The background from the electroweak process is determined using the simulation.
Drell-Yan, $W\to \tau\nu$, Diboson (WW/WZ/ZZ), and $t\bar{t}$ processes are considered 
for the electroweak background.
All backgrounds are estimated for $W^{+}$, $W^{-}$, and $W$, respectively.
Figure \ref{fig:w_met} shows the signal and background distribution in \met for $W\to e\nu$ and
 $W\to \mu\nu$ and Table \ref{tab:wbkg} describes the background rate of each physics process for W boson production.
After background subtraction, we get $136328\pm 386$ of W events in the electron channel 
(the signal yield of $W^{+}\to e^{+}\nu$ = $81568\pm 297$ and $W^{-}\to e^{-}\nu$ = $54760\pm 246$) and 
$140757\pm 383$ of W events in the muon channel 
(the signal yield of $W^{+}\to \mu^{+}\nu$ = $84091\pm 291$ and $W^{-}\to \mu^{-}\nu$ = $56666\pm 240$).

\begin{table}[ht]
  \begin{center}
    \caption{The background rate of physics process in W boson production.}
    \begin{tabular}{|l|c|c|}  \hline
      Background [$\%$] & $W\to e\nu$  & $W\to \mu\nu$ \\ \hline
      Drell-Yan  & 7.6  &  4.6 \\
      $W\to \tau\nu$ & 3.0  & 3.0 \\
      Diboson (WW/WZ/ZZ) & 0.1  & 0.1 \\
      $t\bar{t}$ & 0.4  & 0.4 \\
      Cosmic & $\-$  & $<0.01$ \\
      QCD   &  From fit  & 5.1 \\ \hline
    \end{tabular}
    \label{tab:wbkg}
  \end{center}
\end{table}  

\subsection{Background and Signal Extraction for Z Production}
\begin{figure}[ht]
\centering
\includegraphics[width=80mm]{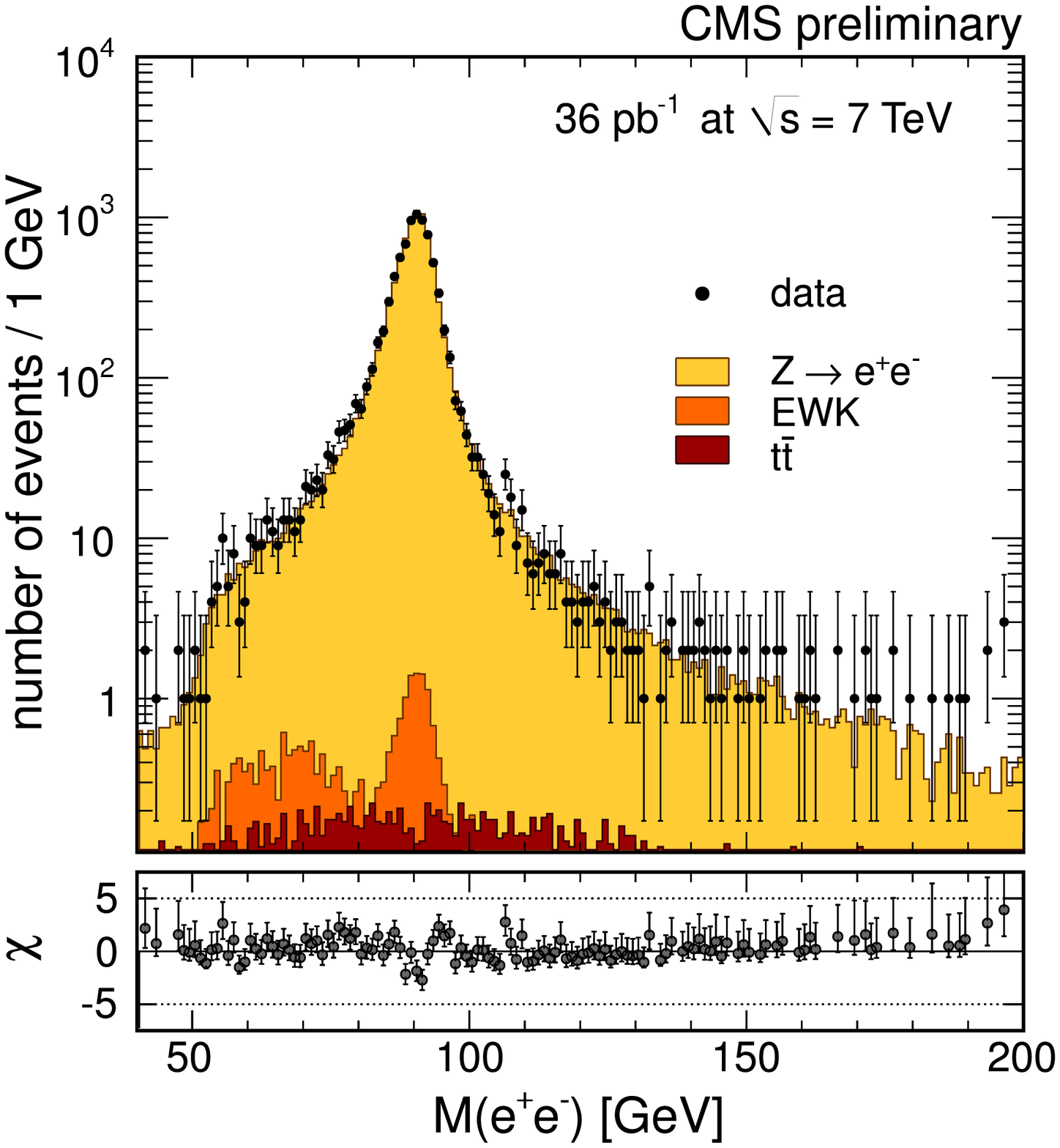}
\includegraphics[width=80mm]{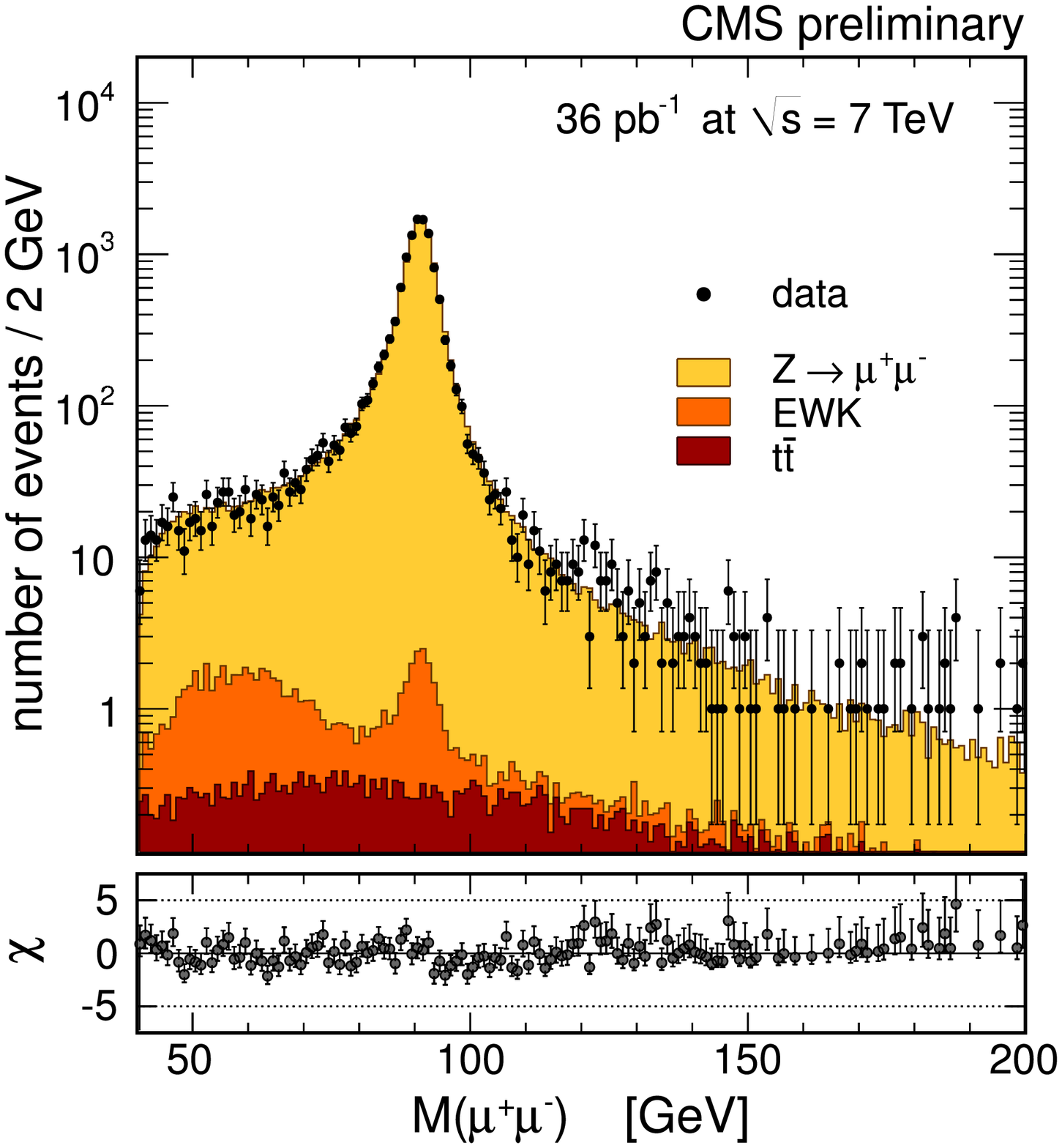}
\caption{Dilepton invariant mass distribution of $Z\to ee$ and $Z\to \mu\mu$ with the background contribution.} \label{fig:zmass}
\end{figure}

The background source of Z boson production is QCD multi-jet and also the electroweak process
 like Diboson, W inclusive, $Z\to \tau\tau$, and $t\bar{t}$ process.
The QCD background is estimated using the isolation fitting method for $Z\to e^{+}e^{-}$
 and the QCD simulation sample is used for $Z\to \mu^{+}\mu^{-}$ because of the low background rate.
The QCD background rate for $Z\to \mu^{+}\mu^{-}$ sample estimated by QCD MC is confirmed by the
 same and opposite charge method in the data and both methods give the consistent background rate.
The electroweak background is estimated using the simulation samples for each physics process
 and the background rates are summarized in Table \ref{tab:zbkg}.
The total background is 0.4 $\%$ level for both $Z\to e^{+}e^{-}$ and $Z\to \mu^{+}\mu^{-}$.
After all background subtraction, we get $8406 \pm 92$ events for $Z\to e^{+}e^{-}$ and
 $13728 \pm 121$ events for $Z\to \mu^{+}\mu^{-}$ process. 
Figure \ref{fig:zmass} shows the dilepton mass distribution with the background contribution.

\begin{table}[ht]
  \begin{center}
    \caption{The background rate of physics process in Z boson production.}
    \begin{tabular}{|l|c|c|}  \hline
      Background [$\%$]       & $Z\to e^{+}e^{-}$  & $Z\to \mu^{+}\mu^{-}$ \\ \hline
      Diboson (WW/WZ/ZZ)      & $0.157\pm 0.001$   & $0.158\pm 0.001$ \\
      $t\bar{t}$              & $0.117\pm 0.008$   & $0.141\pm 0.014$ \\
      $Z\to \tau^{+}\tau^{-}$ & $0.080\pm 0.006$   & $0.124\pm 0.005$ \\
      W+jets                  & $0.010\pm 0.002$   & $0.008\pm 0.002$ \\
      QCD                     & $0.060\pm 0.140$   & $0.013\pm 0.001$ \\ \hline
      Total background        & $0.420\pm 0.140$   & $0.444\pm 0.015$ \\ \hline
    \end{tabular}
    \label{tab:zbkg}
  \end{center}
\end{table}  

%%%%%%%%%%%%%%%%%%%%%%%%%%%%%%%%%%
\section{Systematic Uncertainty}
We consider the systematic uncertainty from the luminosity, efficiencies (reconstruction, ID, and trigger),
 the energy/momentum scale and resolution, \met scale and resolution, the background estimation, 
 PDFs uncertainty for the acceptance, and theoritical uncertainties.  
The largest systematic uncertainty in the measurement is the uncertainty from the integrated luminosity measurement \cite{cmslum}
 which is 4 $\%$. 
Table \ref{tab:syst} summarizes the systematic uncertainty from each systematic source in the measurement.
The total experimental systematic error is comparable with the total theoretical uncertainty.

\begin{table}[ht]
  \begin{center}
    \caption{The systematic uncertainty of the measurement.}
    \begin{tabular}{|l|c|c|c|c|}  \hline
      Source                                    & $W\to e\nu$  & $W\to \mu\nu$ &$Z\to e^{+}e^{-}$  & $Z\to \mu^{+}\mu^{-}$ \\ \hline \hline
      Lepton reconstruction $\&$ identification &  1.4  &  0.9  &  1.8  &  n/a \\
      Trigger prefiring                         &  n/a  &  0.5  &  n/a  &  0.5 \\
      Energy/momentum scale $\&$ resolution     &  0.5  &  0.22 &  0.12 &  0.35 \\
      \met scale $\&$ resolution                &  0.3  &  0.2  &  n/a  &  n/a  \\
      Background subtraction / modeling         &  0.35 &  0.4  &  0.14 &  0.28 \\
      Trigger changes throughout 2010           &  n/a  &  n/a  &  n/a  &  0.1  \\  \hline
      Total experimental                        &  1.6  &  1.1  &  1.8  &  0.7  \\  \hline
      PDF uncertainty for acceptance            &  0.6  &  0.8  &  0.9  &  1.1  \\
      Other theoretical uncertainties           &  0.7  &  0.8  &  1.4  &  1.6  \\  \hline
      Total theoretical uncertainties           &  0.9  &  1.1  &  1.6  &  1.9  \\ \hline
      Total (excluding luminosity)              &  1.8  &  1.6  &  2.4  &  2.0  \\ \hline
    \end{tabular}
    \label{tab:syst}
  \end{center}
\end{table}  

%%%%%%%%%%%%%%%%%%%%%%%%%%%%%%%%%%
\section{W and Z Boson Cross Section Result}
\begin{figure}[ht]
  \centering
  \includegraphics[width=80mm]{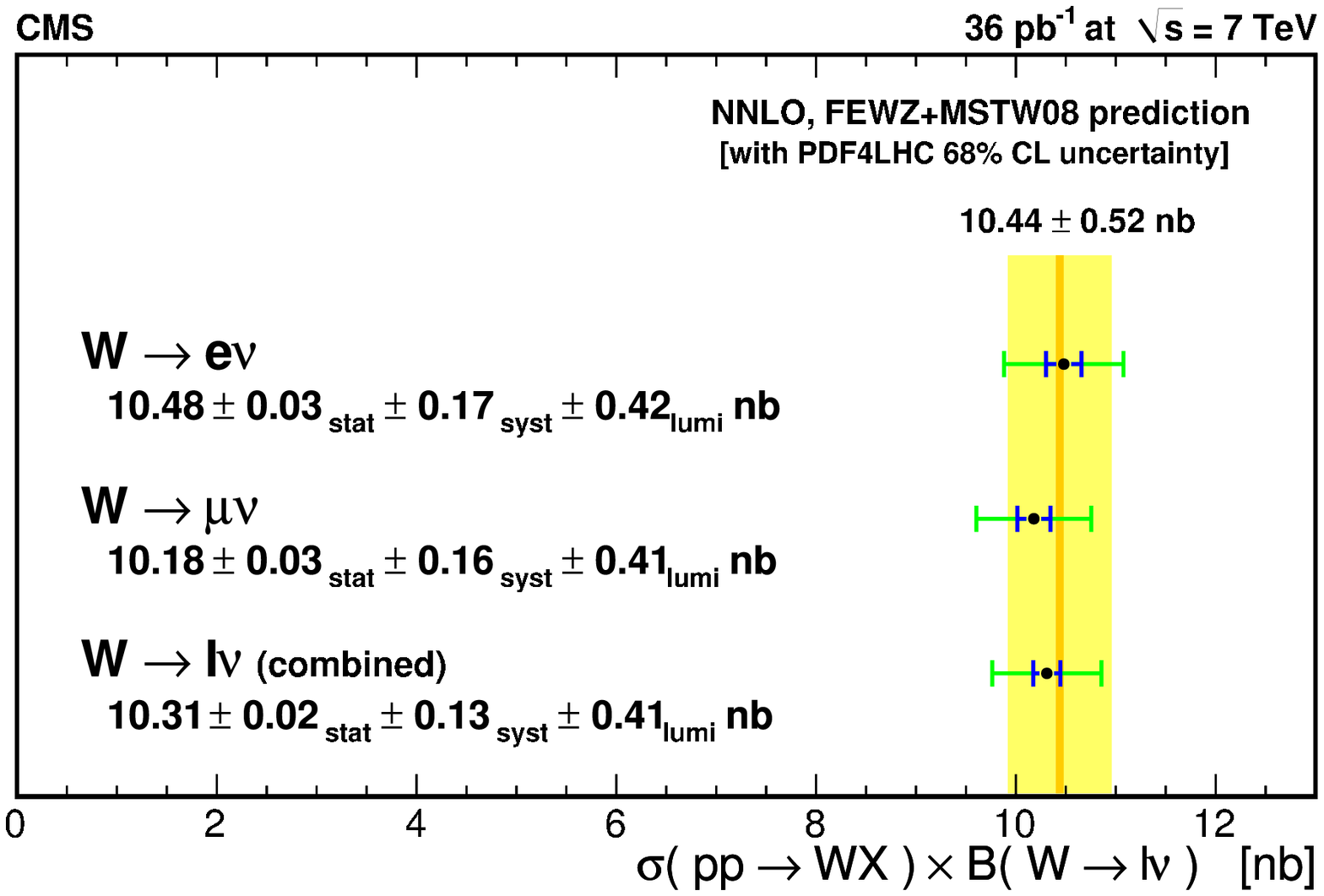}
  \includegraphics[width=80mm]{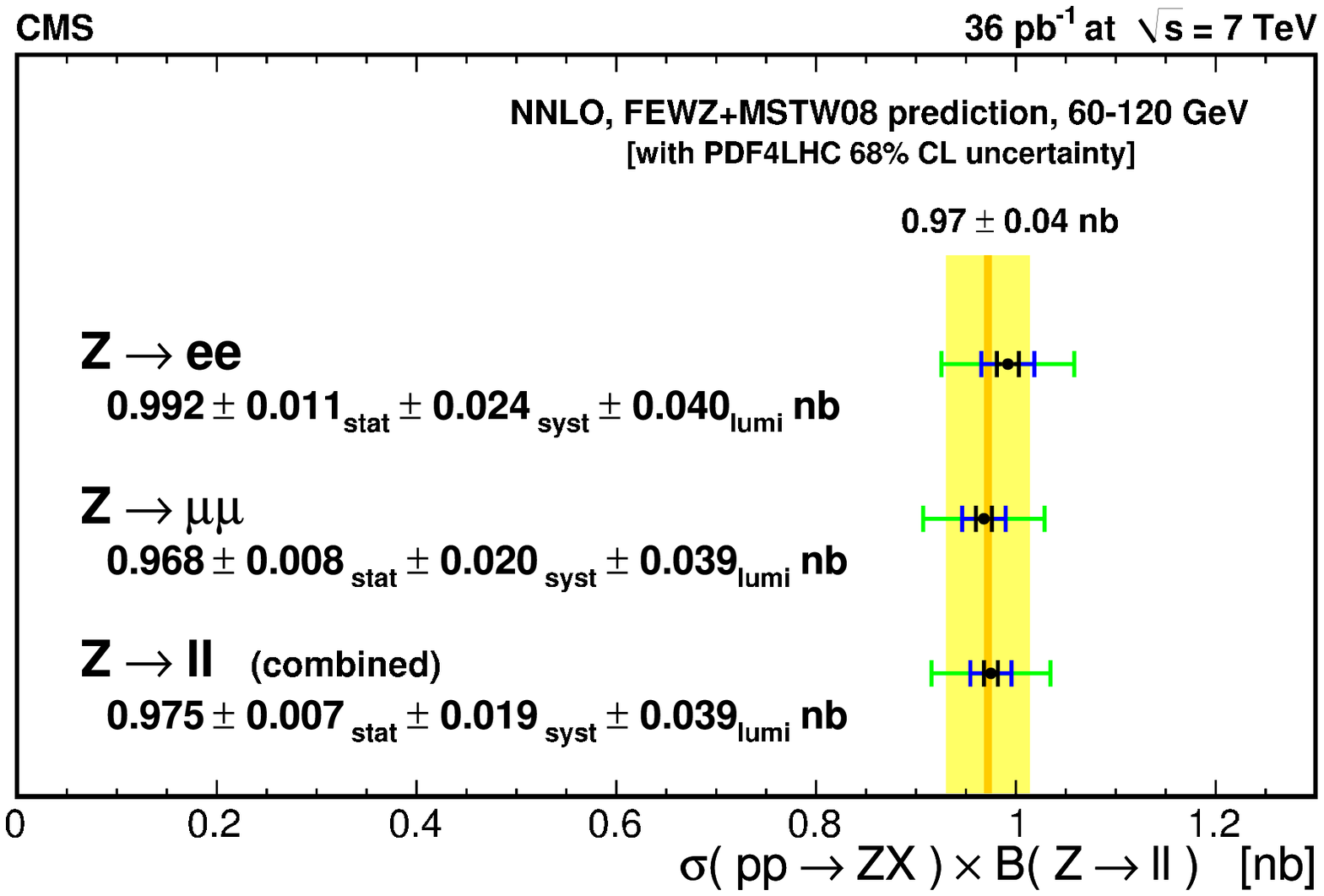}
  \includegraphics[width=80mm]{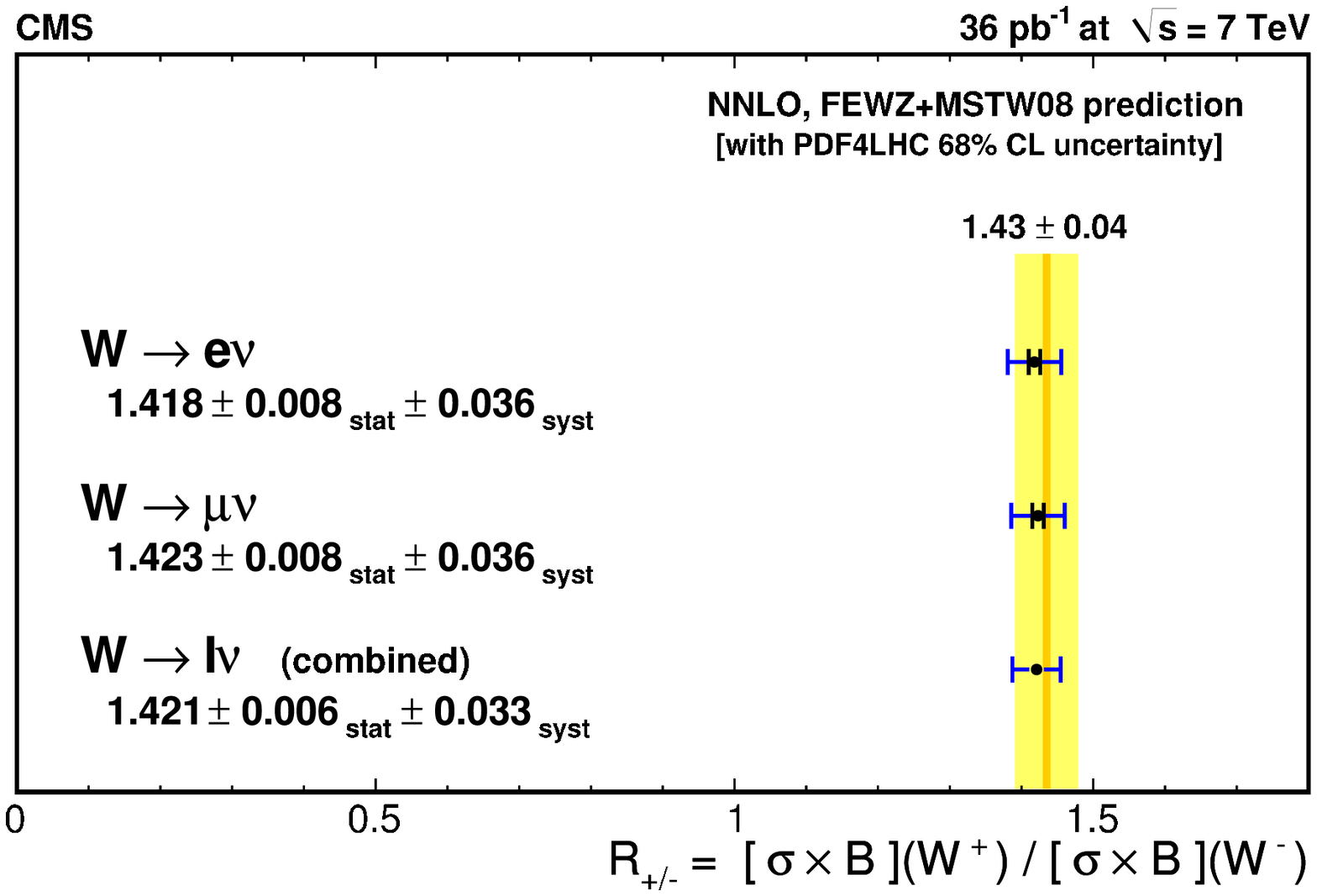}
  \includegraphics[width=80mm]{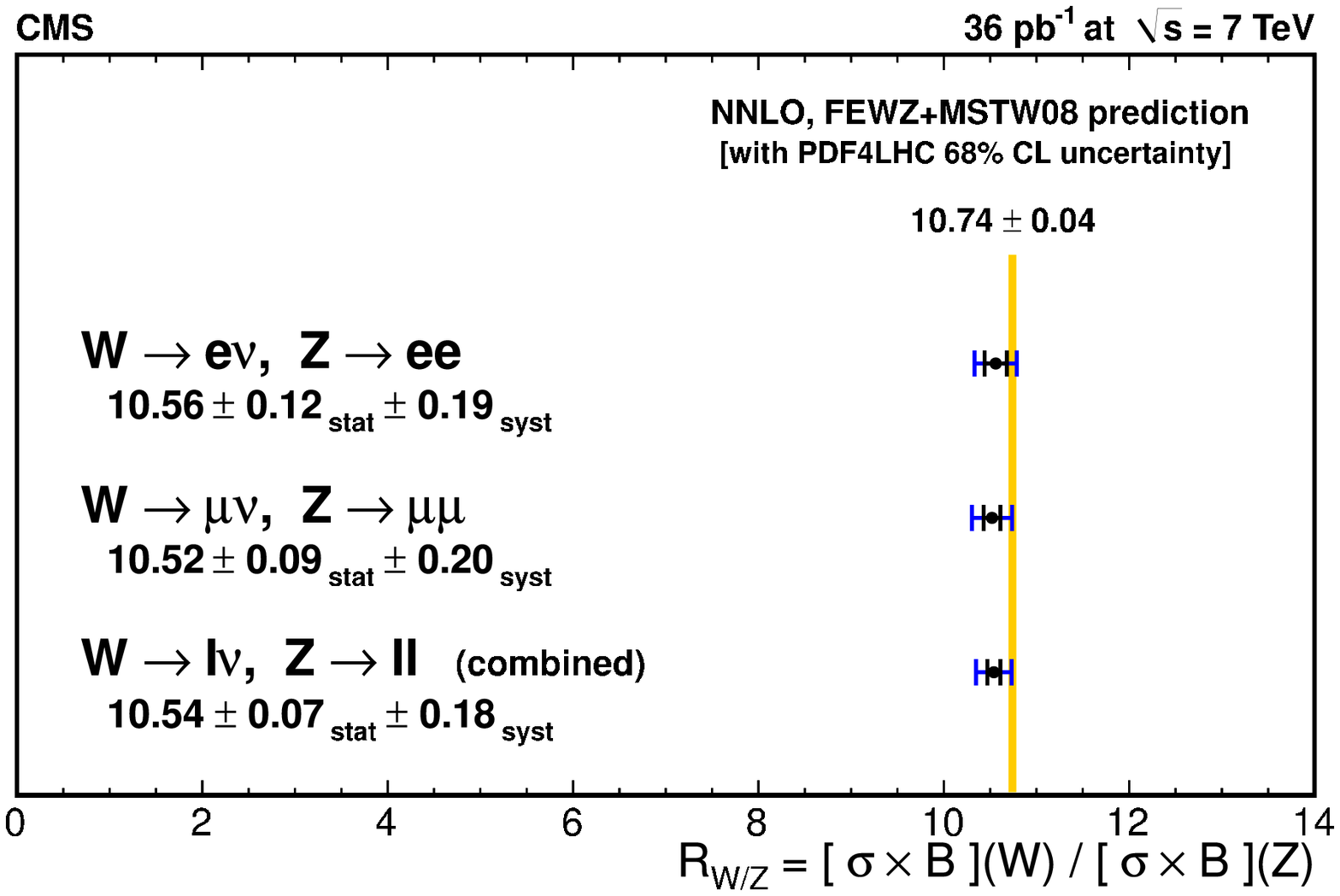}
  \caption{The inclusive cross section of W and Z boson and its ratio.
    The measurements are compared with NNLO theory calculation, \fewz~with \mstw~PDFs.
    The yellow band corresponds to \pdflhc~PDFs uncertainty with 68 $\%$ confidence level.} \label{fig:sigma_sum}
\end{figure}

\begin{figure}[ht]
  \centering
  \includegraphics[width=100mm]{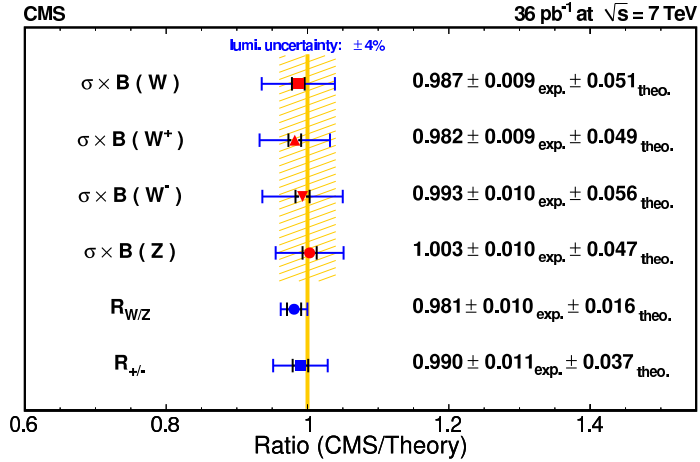}
  \caption{The ratio of the measurement to the theory prediction (NNLO \fewz~with \mstw~PDFs).
    The yellow band corresponds to the luminosity uncertainty, $\pm 4\%$.} \label{fig:data_to_theory}
\end{figure}

The inclusive cross section of W and Z boson is calculated by:
\begin{equation}
\sigma \times B = \frac{(N_{sel} - N_{bkg})}{A\times \epsilon \times \int \mathit{L dt}}
\end{equation}
where $N_{sel}$ is the selected number of events, $N_{bkg}$ is the number of the background events,
 $A\times \epsilon$ is the acceptance times the efficiencies, and $\int \mathit{L dt}$ is the integrated luminosity.
The inclusive cross section is measured for the electron and muon channel, respectively, and then
 these cross sections are combined at the end.
For the ratio of the cross section, we consider the ratio of $W^{+}$ to $W^{-}$ cross section and
 also the ratio of W to Z cross section. The luminosity uncertainty cancels in the ratio of the cross sections.
Table \ref{tab:sigma_sum} summarizes the inclusive cross section of each channel for W and Z boson 
 and also the ratio of $W^{+}$ to $W^{-}$ and W to Z cross section.
The measurements are compared to NNLO theory calculation, \fewz~\cite{fewz1,fewz2} with \mstw~PDFs, which is shown in Figure \ref{fig:sigma_sum}.
The inclusive cross sections and its ratios agree with the prediction within 1 $\sigma$ of total error.
Figure \ref{fig:data_to_theory} shows the ratio of the measurement to the theory prediction (NNLO \fewz).

\begin{table}[ht]
  \begin{center}
    \caption{The inclusive cross section of W and Z boson production and the ratio of $W^{+}$ to $W^{-}$ and W to Z boson.}
    \begin{tabular}{|l|c|}  \hline
      Process         &  Cross Section  \\ \hline
      $W\to e\nu$     &  $10.48\pm 0.03_{stat} \pm 0.17_{syst} \pm 0.42_{lumi}$ nb  \\
      $W\to \mu\nu$   &  $10.18\pm 0.03_{stat} \pm 0.16_{syst} \pm 0.41_{lumi}$ nb  \\  \hline
      $W\to \ell\nu$  &  $10.31\pm 0.02_{stat} \pm 0.13_{syst} \pm 0.41_{lumi}$ nb  \\  \hline \hline
      $Z\to ee$       &  $0.992\pm 0.011_{stat} \pm 0.024_{syst} \pm 0.040_{lumi}$ nb  \\
      $Z\to \mu\mu$   &  $0.968\pm 0.008_{stat} \pm 0.020_{syst} \pm 0.039_{lumi}$ nb  \\  \hline
      $Z\to \ell\ell$ &  $0.975\pm 0.007_{stat} \pm 0.019_{syst} \pm 0.039_{lumi}$ nb  \\  \hline 
    \end{tabular}
    \begin{tabular}{|l|c|}  \hline
      Process  &  Cross Section Ratio  \\ \hline
      $W^{+}/W^{-}(W\to e\nu)$      & $1.418\pm 0.008_{stat}\pm 0.036_{syst}$  \\
      $W^{+}/W^{-}(W\to \mu\nu)$    & $1.423\pm 0.008_{stat}\pm 0.036_{syst}$  \\  \hline
      $W^{+}/W^{-}(W\to \ell\nu)$   & $1.421\pm 0.006_{stat}\pm 0.033_{syst}$  \\  \hline \hline
      $W\to e\nu/Z\to ee$           & $10.56\pm 0.12_{stat}\pm 0.19_{syst}$  \\
      $W\to \mu\nu/Z\to \mu\mu$     & $10.52\pm 0.09_{stat}\pm 0.20_{syst}$  \\  \hline
      $W\to \ell\nu/Z\to \ell\ell$  & $10.54\pm 0.07_{stat}\pm 0.18_{syst}$  \\  \hline
    \end{tabular}
    \label{tab:sigma_sum}
  \end{center}
\end{table}  

%%%%%%%%%%%%%%%%%%%%%%%%%%%%%%%%%%
\section{Lepton Charge Asymmetry of W Production}
In W production, it is hard to measure the asymmetry of W production directly because the momentum of $\nu$ in the longitudinal
 direction is unknown. Therefore, the asymmetry is measured using the final state lepton, which is the lepton charge asymmetry.
The lepton charge asymmetry of W production ($A(\eta)$) is defined as:
\begin{equation}
A(\eta) = \frac{d\sigma/d\eta(W^{+}\to \ell^{+}\nu) - d\sigma/d\eta(W^{-}\to \ell^{-}\bar{\nu})}{d\sigma/d\eta(W^{+}\to \ell^{+}\nu) + d\sigma/d\eta(W^{-}\to \ell^{-}\bar{\nu})}
\end{equation}
where $\eta$ is the pseudorapidity of the lepton in the final state.
The lepton charge asymmetry has a strong $\eta$ dependence, so it is important to understand $\eta$ dependence of
 efficiencies and background.
In the measurement, the efficiencies (reconstruction, ID, and trigger) are estimated as a function of muon $p_{\rm T}$ and $\eta$.
The background is also estimated in $\eta$ using the shape fitting method (\met fit for the electron channel
 and the isolation energy fitting for the muon channel) from the data directly.
For the electroweak background, we use the simulation samples.
All other methods of the measurement are same as used for the inclusive cross section measurement.

\begin{figure}[ht]
  \centering
  \includegraphics[width=100mm]{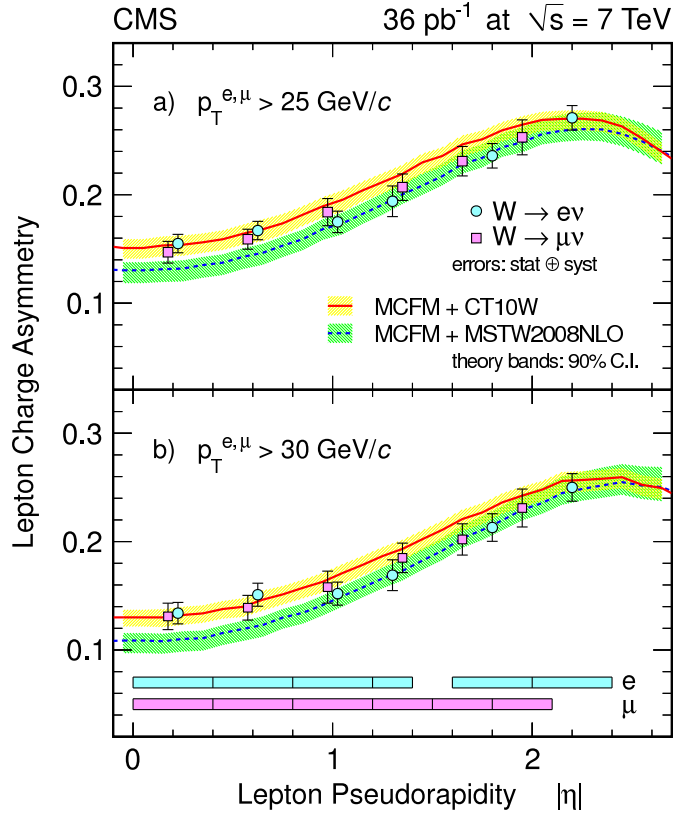}
  \caption{The lepton charge asymmetry of W production.
    The measurement is compared to the theory prediction, \mcfm~with CTEQ10W PDFs and NLO MSTW2008 PDFs.} \label{fig:asym}
\end{figure}

The lepton charge asymmetry has $p_{\rm T}$ dependence, so it is measured in two muon $p_{\rm T}$ bins, $p_{\rm T}(\mu)>25$ and $30$ GeV.
The asymmetry is measured in the electron and muon channel, respectively, and shows a good agreement between these two channels.
Figure \ref{fig:asym} shows the lepton charge asymmetry in $\eta$ and $p_{\rm T}$ compared to the theory predictions,
 \mcfm~\cite{mcfm} with CTEQ10W \cite{cteq} and NLO MSTW2008 \cite{mstw} PDFs.
The precision of the measurement is less than 1.1 $\%$ for the statistical uncertainty and 1.5 $\%$ for the total uncertainty in all
 $\eta$ bins.
This measurement is new input to PDF global fit which constrains to the u and d quark ratio.

%%%%%%%%%%%%%%%%%%%%%%%%%%%%%%%%%%
\section{Conclusion}
We measure the inclusive cross section of W and Z boson production and its ratio using
 36 $pb^{-1}$ data at CMS.
The inclusive cross section measurements show a good agreement with NNLO QCD theory prediction.
The lepton charge asymmetry of W production is also measured and compared with various PDFs set, CTEQ10W and NLO MSTW2008.
More analysis detail is in Ref. \cite{cmspas1} for the inclusive cross section measurement and Ref. \cite{cmspas2} for
 the lepton charge asymmetry measurement.

% If you have acknowledgments, this puts in the proper section head.
%\bigskip % extra skip inserted
%%%%%%%%%%%%%%%%%%%%%%%%%%%%%%%%%%
\begin{acknowledgments}
         We thank the technical and administrative staff at CERN and other CMS institutes,
and acknowledge support from: FMSR (Austria); FNRS and FWO (Belgium); CNPq, CAPES,
FAPERJ, and FAPESP (Brazil); MES (Bulgaria); CERN; CAS, MoST, and NSFC (China); COLCIENCIAS (Colombia);
 MSES (Croatia); RPF (Cyprus); Academy of Sciences and NICPB (Estonia);
 Academy of Finland, ME, and HIP (Finland); CEA and CNRS/IN2P3 (France); BMBF,
DFG, and HGF (Germany); GSRT (Greece); OTKA and NKTH (Hungary); DAE and DST (India);
 IPM (Iran); SFI (Ireland); INFN (Italy); NRF and WCU (Korea); LAS (Lithuania);
 CINVESTAV, CONACYT, SEP, and UASLP-FAI (Mexico); PAEC (Pakistan); SCSR (Poland);
 FCT (Portugal); JINR (Armenia, Belarus, Georgia, Ukraine, Uzbekistan); MST and MAE (Russia);
 MSTD (Serbia); MICINN and CPAN (Spain); Swiss Funding Agencies (Switzerland); NSC (Taipei);
UBITAK and TAEK (Turkey); STFC (United Kingdom); DOE and NSF (USA).
\end{acknowledgments}

\bigskip % extra skip inserted
% Create the reference section using BibTeX:
%\bibliography{basename of .bib file}

\begin{thebibliography}{9}   % Use for  1-9  references
%\begin{thebibliography}{99} % Use for 10-99 references

\bibitem{cmsdet} CMS Collaboration, ``The CMS experiment at the CERN LHC'', $JINST$ {\bf 0803:S08004} (2008).

\bibitem{powheg1} S. Alioli, P. Nason, C. Oleari et al., ``NLO vector-boson production matched with shower in POWHEG'', $JHEP$ {\bf 07} (2008) 060,
 arXiv:0805.4802. doi:10.1088/1126-6708/2008/07/060.

\bibitem{powheg2} P. Nason, ``A new method for combining NLO QCD with shower Monte Carlo algorithms'', $JHEP$ {\bf 11} (2004) 040,
 arXiv:hep-ph/0409146. doi:10.1088/1126-6708/2004/11/040.

\bibitem{powheg3} S. Frixione, P. Nason, and C. Oleari, ``Matching NLO QCD computations with Parton Shower simulations: the POWHEG method'',
$JHEP$ {\bf11} (2007) 070, arXiv:0709.2092. doi:10.1088/1126-6708/2007/11/070.

\bibitem{geant} GEANT: CERN Program Library Long Writeup W5013.

\bibitem{cmslum} CMS Collaboration, ``Measurement of CMS luminosity'', CMS PAS {\bf EWK-2010-004} (2010).

\bibitem{fewz1} K. Melnikov and F. Petriello, ``Electroweak gauge boson production at hadron colliders through O(alpha(s)**2)'', 
$Phys. Rev.$ {\bf D74} (2006) 114017, arXiv:hep-ph/0609070. doi:10.1103/PhysRevD.74.114017.

\bibitem{fewz2} K. Melnikov and F. Petriello, ``The W boson production cross section at the LHC through O($\alpha_{s}^{2}$)'',
$Phys. Rev. Lett.$ {\bf 96} (2006) 231803, arXiv:hep-ph/0603182. doi:10.1103/PhysRevLett.96.231803.

\bibitem{mcfm} J. M. Campbell and R. K. Ellis, ``Radiative corrections to Z b anti-b production'',
 $Phys. Rev.$ {\bf D62} (2000) 114012, arXiv:hep-ph/0006304. doi:10.1103/PhysRevD.62.114012.

\bibitem{cteq} H.-L. Lai et al., ``New parton distributions for collider physics'', $Phys. Rev.$ {\bf D82} (2010) 074024,
 arXiv:1007.2241. doi:10.1103/PhysRevD.82.074024.

\bibitem{mstw} A. D. Martin, W. J. Stirling, R. S. Thorne et al., ``Parton distributions for the LHC'',
 $Eur. Phys. J.$ {\bf C63} (2009) 189-285, arXiv:0901.0002. doi:10.1140/epjc/s10052-009-1072-5.

\bibitem{cmspas1} CMS Collaboration, ``Measurement of Inclusive W and Z Cross Sections in pp Collisions at $\sqrt{s}$ = 7 TeV'',
 CMS PAS {\bf EWK-2010-005} (2010), arXiv:1107.4789.

\bibitem{cmspas2} CMS Collaboration, ``Measurement of the lepton charge asymmetry in inclusive W production in pp Collisions at $\sqrt{s}$ = 7 TeV'',
 CMS PAS {\bf EWK-2010-006} (2010), $JHEP$ {\bf1104} (2011) 050, arXiv:1103.3470, doi:10.1007/JHEP04(2011)050.

\end{thebibliography}

\end{document}